# Joint Transmit and Receive Antenna Selection System for MIMO-NOMA with Energy Harvesting

Mahmoud Aldababsa and Ertugrul Basar, Senior Member, IEEE [1]

*Abstract*—In this paper, outage probability (OP) of a joint transmit and receive antenna selection (JTRAS) scheme is analyzed in multiple-input multiple-output non-orthogonal multiple access based downlink energy harvesting (EH) relaying networks. In this dual-hop and amplify-and-forward relaying based network, since the first and second hops are types of single-user and multi-user systems, respectively, the optimal JTRAS and suboptimal majority-based JTRAS schemes are employed in the first and second hops. The theoretical OP analysis is carried out over Nakagami-*m* fading channels in the cases of perfect and imperfect successive interference cancellation. Finally, Monte Carlo simulations are performed to substantiate the accuracy of the theoretical analysis. It is shown that the optimal power splitting ratios at the EH relay are different for users and the users with good channel conditions have minimum optimal ratios.

## 1 Introduction

Non-orthogonal multiple access (NOMA) is one of the promising multiple-access candidates for beyond 5G wireless networks [1]. It has potential advantages in terms of spectral efficiency and massive connectivity. The distinctive feature of NOMA is that it serves multiple users simultaneously by allocating the same non-orthogonal resources and different power levels utilizing the superposition coding (SC) scheme at the transmitter while uses the successive interference cancellation (SIC) principle to separate the superimposed signal at the receivers [2].

By extending NOMA to multi-antenna and cooperative communication systems, it attains better outage probability (OP) and ergodic rate compared to the conventional orthogonal multiple access (OMA) [3]-[9]. Specifically, in [3], the OP performance of the NOMA based relaying network is investigated over Rayleigh fading channels. In this network, a multi-antenna base station (BS) communicates with multi-antenna users through a channel state information (CSI) based gain amplify-and-forward (AF) single-antenna relay. Also, the transmit antenna selection (TAS) and maximal ratio combining (MRC) schemes are applied at the BS and users, respectively. Then, the work [3] is extended to [4]-[9] for the fixed-gain AF relay with perfect CSI (pCSI) [4], for the CSI based gain AF relay with imperfect CSI (ipCSI) [5], and for both CSI based and fixed-gain AF relay in the ipCSI and imperfect successive interference cancellation (ipSIC) cases [6]. The OP and ergodic sum-rate are studied in [7] under Rayleigh fading channels for the dual-hop NOMA network, where joint transmit receive antenna selection (JTRAS) and receive antenna selection (RAS) are applied in the first and second hops, respectively. The OP performance of different antenna selection schemes is investigated for dual-hop NOMA network with CSI based AF relay [8]-[9]. Particularly, in [8], the MRT/RAS technique is applied in both hops and the OP performance is investigated with ipCSI. In [9], the optimal schemes (TAS/MRC or JTRAS) and suboptimal (majority-based TAS/MRC or JTRAS) are employed in the first and second hops, respectively and the OP is examined in the presence of ipCSI and feedback delay.

---
[1] Mahmoud Aldababsa and Ertugrul Basar are with the Communications Research and Innovation Laboratory (CoreLab), Department of Electrical and Electronics Engineering, Koc University, Sariyer 34450, Istanbul, Turkey. (e-mail: maldababsa@ku.edu.tr, ebasar@ku.edu.tr)

It is noteworthy to mention that the previous wireless cooperative networks [3]-[9] do not take into account that the relay may have limited battery reserves and may need to rely on some external charging mechanism in order to remain active in the network. As a result, to cope with this practical problem, energy harvesting (EH) in such networks appears as a particularly important solution as it can enable information relaying. In this context, the study of the cooperative NOMA with wireless EH is carried out in [10]-[24]. Particularly, in [10], the achievable data rate is calculated for a two-user NOMA network such that the near user works as an EH relay for the far user. In [11], the ergodic sum-rate performance for the NOMA network with AF EH relay is studied and then its upper bound expression is obtained. The OP expression is derived in closed-form under Nakagami-*m* fading channels for NOMA network with AF EH relay [12]. The impact of the co-channel interferences and hardware impairments on the EH NOMA networks are examined in [12] and [14]-[16], respectively. The EH NOMA networks' performance is investigated through diverse relay selection techniques in [7]-[21]. Extended to the work [3], the OP performance is examined in [22] and [23] for EH AF and decode-and-forward (DF) relays, respectively. In [24], the OP performance is examined over Rayleigh fading channels for the NOMA based EH relaying network, where the schemes JTRAS and RAS are employed in the first and second hops, respectively.

It is worth to mention that [11]-[23] use a relay with single transmit and single receive antenna while [24] uses a relay with single transmit and multiple receive antennas. However, relaying with multi-transmit/receive antennas can offer better system performance than that of the relay with a single antenna. In this context, in order to exploit the advantages of the multiple antennas in all network nodes without extra power consumption and hardware complexity, antenna selection is expected to be employed in both hops. However, since the optimal JTRAS (JTRAS-opt) scheme is not possibly achieved in the multi-user NOMA networks (as in the second hop)[2] and may be achieved in the single-user communication network (as in the first hop), the JTRAS-opt and suboptimal majority-based JTRAS (JTRAS-maj)[3] schemes are employed in the first and second hops, respectively.

To the best of our knowledge, there has not been any attempt to apply MIMO-NOMA into the EH relaying networks in the literature, which highly motivates this paper. Furthermore, the SIC error (ipSIC) is also a practical and strong problem that NOMA networks encounter. The key contributions of this paper can be highlighted as follows:
- The OP performance is examined for a cooperative MIMO-NOMA networks in both pSIC and ipSIC cases. In this network, the BS, EH relay and users have multiple antennas and the JTRAS-opt and JTRAS-maj schemes are applied in the first and second hops, respectively.
- In the cases of pSIC and ipSIC, the exact OP expression is derived in closed-form over the generic channel model identical and independent distributed (i.i.d.) Nakagami-$m$ fading which also includes the special channel model line-of-sight (LoS) when $m = 1$, i.e., Rayleigh fading.
- Finally, the theoretical results are validated by the Monte Carlo simulation. The results show that the optimum power division ratios by the power splitter at the EH relay are different for users and the users with good channel conditions will have less transmission power.

---

[2] This is because, in the NOMA network, all the users have to be taken into account in order to find the best transmit-receive antenna pair, which is for all the users may not be possible for all channel realizations. As well, due to severe inter-user interference, mathematical analyses of the antenna selection based NOMA networks will be more difficult because of using SC and SIC methods at the transmitter and receiver, respectively.

[3] There are also another suboptimal antenna selection solutions such as max-max-max and max-min-max [25]-[27]. However, the results in [9], [28]-[31] show the superiority of JTRAS-maj scheme over the max-max-max and max-min-max based JTRAS schemes. Because of that, in this paper, we attend to use the JTRAS-maj scheme.

The numerical results also demonstrate that although ipSIC affects OP performance, it does not affect the optimal power splitting ratios. Furthermore, the optimal EH relay location has to be closer to the BS and as the EH relay becomes near the users, the OP becomes worse and converges to the same value for all users.

The rest of this paper is organized as follows. In Section 2, the network model, antenna selection schemes and signal-to-interference-and-noise ratio statistics are introduced in detail. In Section 3, the OP analysis is conducted. At last, the numerical results are presented to verify theoretical OP analysis in Section 4 and followed by concluding remarks in Section 5.

## 2 Network Model

In Fig. 1, we consider a cooperative MIMO-NOMA network in which a BS communicates with $K$ users with the help of an AF EH-relay. All nodes have multiple antennas such that $N_S$ transmit and $N_U$ receive antennas are employed at the BS and each user, respectively as well as $N_{Rt}$ transmit and $N_{Rr}$ receive antennas are equipped at the relay. We assume that the direct links between the BS and users are blocked. The channel coefficients corresponding to the BS-relay and relay-$U_k$ links are denoted by $\{h_{SR}^{(i,j)}, 1 \leq i \leq N_S, 1 \leq j \leq N_{Rr}\}$ and $\{h_{RU_k}^{(i,j)}, 1 \leq i \leq N_{Rt}, 1 \leq j \leq N_U\}$, respectively, where $i$ and $j$ refer, respectively to the $i$th transmit and $j$th receive antennas corresponding to each communication node (BS, relay and users).

The whole communication process between the BS and users is accomplished in two stages. At the first stage, the BS transmits $x_S = \sqrt{P_S a_k} s_k + \sum_{t \neq k}^{K} \sqrt{P_S a_t} s_t$ to the relay, where $P_S$, $a_t$ and $s_t$ are the transmit power at the BS, power factor and information of user $t$, respectively. Based on NOMA, the power coefficients have two features; $a_1 + \ldots + a_K = 1$ and reversely ordered according to the second hop channel gains as $a_1 \geq \ldots \geq a_K$. If $h_{SR}$ is the channel coefficient corresponding to the best link (the red link as shown in Fig. 1) from BS-relay links, offering the highest channel gain between the BS and the relay, and $n_R$ is the zero mean complex additive Gaussian noise with a variance of $\sigma_R^2$ at the relay, then the received signal at the relay can be stated as

$$y_R = h_{SR}x_S + n_R$$
$$= h_{SR}\sqrt{P_S a_k}s_k + h_{SR}\sum_{t\neq k}^{K}\sqrt{P_S a_t}s_t + n_R. \qquad (1)$$

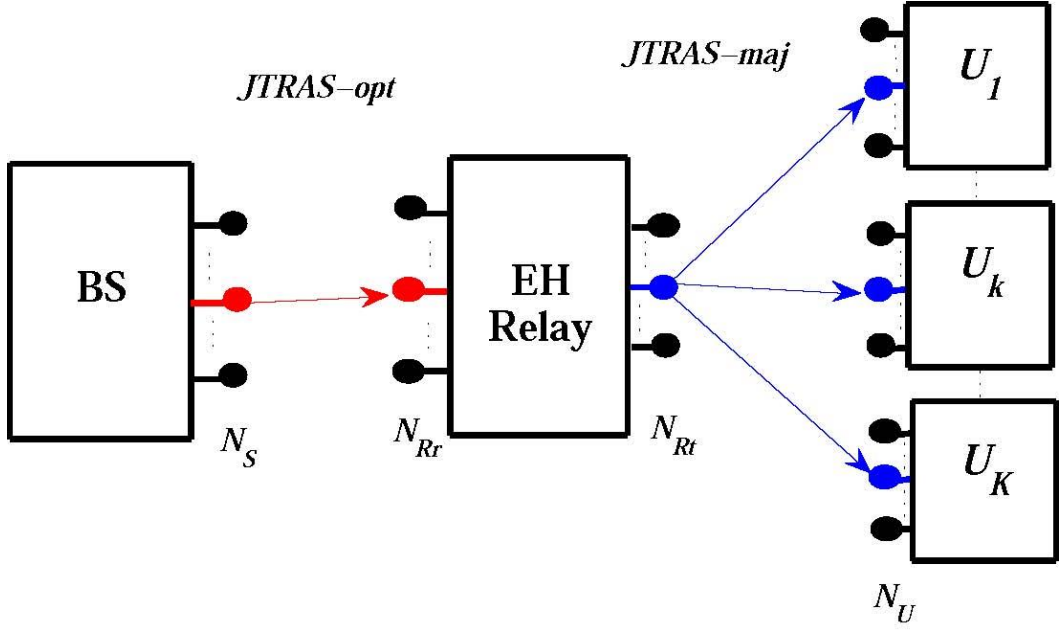

Figure 1: Network model of the EH MIMO-NOMA network.

At the second stage, the relay harvests the energy from the signals transmitted by the BS following the power splitting-based relaying (PSR) protocol [32], and then forwards the remaining information to the users. In other words, the received signal $y_R$ is split into two parts with $w:1-w$ proportion by the power splitter. Specifically, $\sqrt{1-w}y_R$ is exploited as information to be processed and $wy_R$ is harvested as the transmitting power at the relay in order to amplify the remaining $\sqrt{1-w}y_R$ signal with amplification factor $G$ and transmit it to $K$ users. If the transmit power at the relay is

$$P_R = \zeta w|h_{SR}|^2. \qquad (2)$$

Then, the amplification factor can be expressed as

$$G = \sqrt{\frac{P_R}{(1-w)P_S|h_{SR}|^2+\sigma_{SR}^2}}$$
$$= \sqrt{\frac{\zeta w|h_{SR}|^2}{(1-w)P_S|h_{SR}|^2+\sigma_{SR}^2}}$$
$$\approx \sqrt{\frac{\zeta w}{(1-w)}}, \qquad (3)$$

where $0\leq\zeta\leq 1$ is the energy conversion efficiency and $|\cdot|$ represents the absolute value. Accordingly, in the case of ipSIC, the received signal at user $k$ can be expressed as

$$y_{U_k} = \underbrace{h_{RU_k}G\sqrt{1-w}h_{SR}\xi\sum_{t=1}^{k-1}a_t}_{\text{ipSIC term}}$$
$$+ \underbrace{h_{RU_k}G\sqrt{1-w}h_{SR}\sqrt{P_S a_k}s_k}_{\text{Desired signal term}}$$
$$+ \underbrace{h_{RU_k}G\sqrt{1-w}h_{SR}\sum_{t=k+1}^{K}\sqrt{P_S a_t}s_t}_{\text{Interference term}}$$
$$+ \underbrace{h_{RU_k}Gn_R + n_{U_k}}_{\text{Noise term}}, \qquad (4)$$

where $h_{U_k}$ is the channel coefficient corresponding to relay-users links between the majority transmit antenna at the relay and the best receive antennas at the users (the blue links as shown in Fig. 1)[4], $n_{U_k}$ is the zero mean complex additive Gaussian noise with a variance of $\sigma_{U_k}^2$ at user $k$ and $\xi$ is the error propagation factor caused as a result of ipSIC process. In this paper, we use the ipSIC model that followed Gaussian distribution with a variance of $0 \leq \xi \leq 1$ [33]. It is worth mentioning that the received signal at user $k$ in the case of the pSIC can be simply expressed by putting $\xi = 0$ into (4).

## 2.1 Antenna selection schemes

As shown in Fig. 1, the JTRAS-opt [34] and suboptimal JTRAS-maj [9], [28]-[31] schemes are applied in the first and second hops, respectively. In particular, by applying the JTRAS-opt, the transmit-receive antenna pair between the BS and the relay, $(i_S, j_R)$ is chosen, where $i_S \in \{1, \ldots, N_S\}$ and $j_R \in \{1, \ldots, N_{Rr}\}$. Precisely, $(i_S, j_R)$ has the highest channel gain between the transmit antennas of BS and receiver antennas of the relay and it can be expressed as

$$(i_S, j_R) = \arg\max_{\substack{1 \leq i \leq N_S \\ 1 \leq j \leq N_{Rr}}}\{|h_{SR}^{(i,j)}|^2\}. \qquad (5)$$

On the other hand, the JTRAS-maj scheme can be realized in the second hop by first applying a JTRAS-opt [34] separately for the users. Therefore, the transmit-receive antenna pair, $(i_{RU_k}, j_{U_k})$, providing the highest channel gain between the BS and user $k$ is determined, where $i_{RU_k} \in \{1, \ldots, N_{Rt}\}$ and $j_{U_k} \in \{1, \ldots, N_U\}$. Then, by the majority algorithm [9], [28]-[31], the majority transmit antenna at the relay $i_R \in \{1, \ldots, N_{Rt}\}$ is specified. Accordingly, the $(i_{RU_k}, j_{U_k})$ and $i_R$ can be stated, respectively as

$$(i_{RU_k}, j_{U_k}) = \arg\max_{\substack{1 \leq i \leq N_{Rt} \\ 1 \leq j \leq N_U}}\{|h_{RU_k}^{(i,j)}|^2\}, k = 1, \ldots, K. \qquad (6)$$

$$i_R = Maj(i_{RU_k}). \qquad (7)$$

In (7), $Maj(\cdot)$ denotes the majority function that determines the majority transmit antenna through an algorithm given in detail in [30].

---

[4] Recall, based on the NOMA concept, the power coefficients are ordered reversely to the second hop channel gains, i.e., $a_1 \geq \ldots \geq a_K$ opposite to $|h_{U_1}|^2 \leq \ldots \leq |h_{U_K}|^2$

## 2.2 Signal-to-interference-and-noise ratio statistics

From (4), the expression of the instantaneous signal-to-interference-and-noise ratio (SINR) of the $U_k$ to detect the $U_l$ can be derived in the case of ipSIC as

$$\gamma_{l,k} = \frac{\bar{\gamma}|h_{SR}|^2|h_{RU_k}|^2 a_l}{\bar{\gamma}|h_{SR}|^2|h_{RU_k}|^2\Sigma_l + c_1|h_{RU_k}|^2 + c_2}, l \neq K, l < k. \tag{8}$$

In (8), $\bar{\gamma} = \frac{P_S}{\sigma^2}$ is the SNR, $\sigma^2 = \sigma_R^2 = \sigma_{U_k}^2$, $\Sigma_l = \xi \sum_{t=1}^{l-1} a_t + \sum_{t=l+1}^{K} a_t$, $c_1 = \frac{1}{1-w}$ and $c_2 = \frac{1}{\zeta w}$.

In this paper, we assume that the magnitudes of the fading gains are i.i.d. with Nakagami-$m$ distribution. Thus, the cumulative distribution function (CDF) and the probability density function (PDF) of the square of Nakagami-$m$ random variable $X$ are stated, respectively as

$$F_X(x) = \frac{\psi(m, \frac{mx}{\Omega})}{\Gamma(m)}. \tag{9}$$

$$f_X(x) = \left(\frac{m}{\Omega}\right)^m \frac{x^{m-1}}{\Gamma(m)} e^{-\frac{mx}{\Omega}}. \tag{10}$$

Note that $\psi(\cdot,\cdot)$ and $\Gamma(\cdot)$ in (9)-(10) are lower incomplete Gamma and Gamma functions, respectively, $m$ is the parameter of Nakagami-$m$ distribution and $\Omega = E[|X|^2]$, where $E[\cdot]$ refers to expectation operator.

Now, according to the selection criterion in (5), the CDF of the squared channel gain corresponding to the best link in the first hop can be derived by

$$F_{|h_{SR}|^2}(x) = (F_X(x))^{N_S N_{Rr}}. \tag{11}$$

Then, taking the derivative of (11), the PDF of the squared channel gain corresponding to the best link in the first hop can be stated as

$$f_{|h_{SR}|^2}(x) = N_S N_{Rr} f_X(x)(F_X(x))^{N_S N_{Rr}-1}. \tag{12}$$

Using (9), (10) and exploiting the properties of $\psi(\cdot,\cdot)$ in [35], the $F_{|h_{SR}|^2}(x)$ and $f_{|h_{SR}|^2}(x)$ in (11) and (12), respectively can be explicitly rewritten as

$$F_{|h_{SR}|^2}(x) = \left(\frac{\psi\left(m_{SR}, \frac{m_{SR}}{\Omega_{SR}}x\right)}{\Gamma(m_{SR})}\right)^{N_S N_{Rr}}$$

$$= \left(1 - e^{-\frac{m_{SR}}{\Omega_{SR}}x} \sum_{v=0}^{m_{SR}-1} \left(\frac{m_{SR}}{\Omega_{SR}}x\right)^v \frac{1}{v!}\right)^{N_S N_{Rr}}$$

$$= \sum_{u=0}^{N_S N_{Rr}} \sum_{v=0}^{u(m_{SR}-1)} \binom{N_{Rr} N_S}{u} (-1)^u$$

$$\times \vartheta_v(u, m_{SR}) x^v e^{-\frac{u m_{SR}}{\Omega_{SR}}x}. \tag{13}$$

$$f_{|h_{SR}|^2}(x) = N_S N_{Rr} \left(\frac{m_{SR}}{\Omega_{SR}}\right)^{m_{SR}} \frac{x^{m_{SR}-1}}{\Gamma(m_{SR})} e^{-\frac{m_{SR}}{\Omega_{SR}}x}$$

$$\times \left(1 - e^{-\frac{m_{SR}}{\Omega_{SR}}x} \sum_{n=0}^{m_{SR}-1} \left(\frac{m_{SR}}{\Omega_{SR}}x\right)^n \frac{1}{n!}\right)^{N_S N_{Rr}-1}$$

$$= N_S N_{Rr} \left(\frac{m_{SR}}{\Omega_{SR}}\right)^{m_{SR}} \frac{x^{m_{SR}-1}}{\Gamma(m_{SR})} e^{-\frac{m_{SR}}{\Omega_{SR}}x}$$

$$\times \sum_{u=0}^{N_S N_{Rr}-1} \sum_{v=0}^{u(m_{SR}-1)} \binom{N_S N_{Rr}-1}{u} (-1)^u$$

$$\times \vartheta_v(u, m_{SR}) x^v e^{-\frac{u m_{SR}}{\Omega_{SR}}x}. \tag{14}$$

Note that in (13)-(14), $\vartheta_x(y, g_z)$ denotes multinomial coefficients which can be defined as $\vartheta_x(y, g_z) = \frac{1}{xd_0}\sum_{o=1}^{x}(w(y+1)-x)d_o\vartheta_{x-y}(y, g_z)$, $x \geq 1$ [35, Eq.(0.314)]. Here, $d_o = (g_z/\Omega_z)^o/o!$, $\vartheta_0(y, g_z) = 1$, and $\vartheta_x(y, g_z) = 0$ if $o > g_z - 1$.

On the other hand, according to the selection criteria in (6)-(7), the CDF of the squared channel gain corresponding to relay-$U_k$ link between the majority transmit antenna at the relay and the best receive antenna at the user $k$ i.e., $F_{|h_{RU_k}|^2}(x)$ is given generally for $K$ users in [30, Eq.(15)]. Nevertheless, in order to avoid majority function's complexity [9], [28]-[31], in this paper, we assume $N_S = 2$ and $K = 3$. So, using [30, Eq.(14)], the CDF of the $F_{|h_{RU_k}|^2}(x)$ is given as

$$F_{|h_{RU_k}|^2}(x) = \sum_{q=1}^{3N_{Rt}} \eta_{(k,q)} \left(\frac{\psi(m_{RU}, \frac{m_{RU}}{\Omega_{RU}}x)}{\Gamma(m_{RU})}\right)^{qN_U}$$

$$= \sum_{q=1}^{3N_{Rt}} \eta_{(k,q)}$$

$$\times \left(1 - e^{-\frac{m_{RU}}{\Omega_{RU}}x} \sum_{S=0}^{m_{RU}-1} \left(\frac{m_{RU}}{\Omega_{RU}}x\right)^S \frac{1}{S!}\right)^{qN_U}$$

$$= \sum_{q=1}^{3N_{Rt}} \sum_{p=0}^{qN_U} \sum_{S=0}^{p(m_{RU}-1)} \binom{qN_U}{p} (-1)^p \eta_{(k,q)}$$

$$\times \vartheta_S(p, m_{RU}) x^S e^{-\frac{p m_{RU}}{\Omega_{RU}}x}, \tag{15}$$

where $\eta_{(1,1)} = \eta_{(1,2)} = \frac{3}{2}, \eta_{(1,3)} = -\frac{9}{2}, \eta_{(1,4)} = \frac{15}{4}, \eta_{(1,5)} = -\frac{3}{2}, \eta_{(1,6)} = \frac{1}{4}, \eta_{(2,3)} = 3, \eta_{(2,4)} = -\frac{3}{4}, \eta_{(2,5)} = -\frac{9}{4}, \eta_{(2,6)} = 1, \eta_{(3,5)} = \frac{3}{2}, \eta_{(3,6)} = -\frac{1}{2}$ and otherwise equal zero.

## 3 Outage Probability Analysis

The OP of user $k$ is defined as the probability that instantaneous SINR of user $k$ to detect user $l$ ($\gamma_{l,k}$) is below a threshold value ($\gamma_{th_l}$), i.e., $OP_k = P_r(\gamma_{l,k} < \gamma_{th_l})$. Accordingly, by using (8), the $OP_k$ can be derived as

$$OP_k = P_r\left(|h_{SR}|^2 < \frac{c_1|h_{RU_k}|^2 \gamma_{th_l} + c_2 \gamma_{th_l}}{\overline{\gamma}|h_{RU_k}|^2(a_l - \Sigma_l \gamma_{th_l})}\right)$$

$$= P_r\left(|h_{RU_k}|^2 < \frac{\frac{\gamma_{th_l} c_1}{\overline{\gamma}(a_l - \Sigma_l \gamma_{th_l})} c_2}{c_1\left(|h_{SR}|^2 - \frac{\gamma_{th_l} c_1}{\overline{\gamma}(a_l - \Sigma_l \gamma_{th_l})}\right)}\right). \tag{16}$$

Now, assume $\tau_k^* = \max\left\{\frac{\gamma_{th_l}c_1}{\overline{\gamma}(a_l - \Sigma_l\gamma_{th_l})}\right\}_{l=1,\ldots,k}$, and under the condition $a_l - \Sigma_l\gamma_{th_l} > 0$, the $OP_k$ can be rewritten as

$$\begin{aligned}
OP_k &= P_r\left(|h_{RU_k}|^2 < \frac{\tau_k^* c_2}{c_1(|h_{SR}|^2 - \tau_k^*)}, |h_{SR}|^2 > \tau_k^*\right) \\
&= 1 - P_r\left(|h_{RU_k}|^2 > \frac{\tau_k^* c_2}{c_1(|h_{SR}|^2 - \tau_k^*)}, |h_{SR}|^2 > \tau_k^*\right) \\
&= 1 - \int_{\tau_k^*}^{\infty}\left(1 - F_{|h_{RU_k}|^2}\left(\frac{\tau_k^* c_2}{c_1(x - \tau_k^*)}\right)\right)f_{|h_{SR}|^2}(x)dx \\
&= 1 - \int_{\tau_k^*}^{\infty} f_{|h_{SR}|^2}(x)dx \\
&\quad + \int_{\tau_k^*}^{\infty} F_{|h_{RU_k}|^2}\left(\frac{\tau_k^* c_2}{c_1(x - \tau_k^*)}\right)f_{|h_{SR}|^2}(x)dx \\
&= F_{|h_{SR}|^2}(\tau_k^*) \\
&\quad + \underbrace{\int_{\tau_k^*}^{\infty} F_{|h_{RU_k}|^2}\left(\frac{\tau_k^* c_2}{c_1(x - \tau_k^*)}\right)f_{|h_{SR}|^2}(x)dx}_{I_1}.
\end{aligned} \quad (17)$$

Using (13)-(15), $I_1$ (17) can be rewritten in (18)

$$\begin{aligned}
I_1 &= \sum_{q=1}^{3N_{Rt}} \eta_{(k,q)}\left(1 - F_{|h_{SR}|^2}(\tau_k^*)\right) + \int_{\tau_k^*}^{\infty} F_{|h_{RU_k}|^2}\left(\frac{\tau_k^* c_2}{c_1(x - \tau_k^*)}\right)f_{|h_{SR}|^2}(x)dx \\
&= \sum_{q=1}^{3N_{Rt}} \eta_{(k,q)}\left(1 - F_{|h_{SR}|^2}(\tau_k^*)\right) + \\
&\sum_{q=1}^{3N_{Rt}}\sum_{p=1}^{qN_U}\sum_{s=0}^{p(m_{RU}-1)}\sum_{u=0}^{N_SN_{Rr}-1}\sum_{v=0}^{u(m_{SR}-1)} \binom{qN_U}{p}\binom{N_SN_{Rr}-1}{u}(-1)^{p+u}\eta_{(k,q)}\vartheta_s(p, m_{RU})
\end{aligned}$$

$$\vartheta_v(u, m_{SR})\frac{N_SN_{Rr}}{\Gamma(m_{SR})}\left(\frac{m_{SR}}{\Omega_{SR}}\right)^{m_{SR}} \underbrace{\int_{\tau_k^*}^{\infty}\left(\frac{\tau_k^* c_2}{c_1(x-\tau_k^*)}\right)^s e^{-\frac{pm_{RU}}{\Omega_{RU}}\frac{\tau_k^* c_2}{c_1(x-\tau_k^*)}} x^{m_{SR}-1+v} e^{\left(-\frac{m_{SR}}{\Omega_{SR}} - \frac{um_{SR}}{\Omega_{SR}}\right)x} dx}_{I_2}. \quad (18)$$

Now, by change of variables $y = x - \tau_k^*$, then $I_2$ in (19) can be expressed as

$$\begin{aligned}
I_2 &= \int_0^{\infty} \left(\frac{\tau_k^* c_2}{c_1 y}\right)^s e^{-\frac{pm_{RU}\tau_k^* c_2}{\Omega_{RU} c_1 y}} \underbrace{(y + \tau_k^*)^{m_{SR}-1+v}}_{J} \\
&\times e^{\left(-\frac{m_{SR}}{\Omega_{SR}} - \frac{um_{SR}}{\Omega_{SR}}\right)(y+\tau_k^*)} dy.
\end{aligned} \quad (19)$$

Next, by the help of series expansion, $J$ in (19) can be stated as

$$J = \sum_{z=0}^{m_{SR}-1+v} \binom{m_{SR}-1+v}{z}(\tau_k^*)^{m_{SR}-1+v-z} y^z. \quad (20)$$

Substituting $J$ into (19), $I_2$ can be rewritten as

$$\begin{aligned}
I_2 &= \left(\frac{\tau_k^* c_2}{c_1}\right)^s \sum_{z=0}^{m_{SR}-1+v}\binom{m_{SR}-1+v}{z}(\tau_k^*)^{m_{SR}-1+v-z} \\
&\times e^{\left(-\frac{m_{SR}}{\Omega_{SR}} - \frac{um_{SR}}{\Omega_{SR}}\right)\tau_k^*} \\
&\times \int_0^{\infty} y^{z-s} e^{-\frac{pm_{RU}\tau_k^* c_2}{\Omega_{RU} c_1}y^{-1}} e^{\left(-\frac{m_{SR}}{\Omega_{SR}} - \frac{um_{SR}}{\Omega_{SR}}\right)y} dy.
\end{aligned} \quad (21)$$

Here the integral in (21) can be calculated as $2\left(\frac{b_1}{b_2}\right)^{\frac{z-s+1}{2}} K_{z-s+1}(2\sqrt{b_1 b_2})$ [35, Eq.(3.471.9)], where $b_1 = -\frac{pm_{RU}\tau_k^* c_2}{\Omega_{RU} c_1}$, $b_2 = -\frac{m_{SR}}{\Omega_{SR}} - \frac{um_{SR}}{\Omega_{SR}}$ and $K_\beta(\cdot)$ denotes the $\beta$th order of modified Bessel function of the second kind. Finally, using $I_2$, $I_1$ and (13) in (17), the exact closed-form OP expression in the case of the ipSIC can be expressed in (22).

$$OP_k = \sum_{q=1}^{3N_{Rt}} \eta_{(k,q)}$$

$$+ \left(1 - \sum_{q=1}^{3N_{Rt}} \eta_{(k,q)}\right) \sum_{u=0}^{N_S N_{Rr}} \sum_{v=0}^{u(m_{SR}-1)} \binom{N_S N_{Rr}}{u}(-1)^u \vartheta_v(u, m_{SR})(\tau_k^*)^v e^{-\frac{um_{SR}}{\Omega_{SR}}\tau_k^*}$$

$$+ \sum_{q=1}^{3N_{Rt}} \sum_{p=1}^{qN_U} \sum_{s=0}^{p(m_{RU}-1)} \sum_{u=0}^{N_S N_{Rr}-1} \sum_{v=0}^{u(m_{SR}-1)} \sum_{z=0}^{m_{SR}-1+v} \binom{qN_U}{p}\binom{N_S N_{Rr}-1}{u}\binom{m_{SR}-1+v}{z}$$
$$\times (-1)^{p+u} \eta_{(k,q)} \vartheta_s(p, m_{RU})$$

$$\vartheta_v(u, m_{SR}) \frac{2N_S N_{Rr}}{\Gamma(m_{SR})} \left(\frac{m_{SR}}{\Omega_{SR}}\right)^{m_{SR}} \left(\frac{\tau_k^* c_2}{c_1}\right)^s (\tau_k^*)^{m_{SR}-1+v-z} e^{\left(-\frac{m_{SR}}{\Omega_{SR}} - \frac{um_{SR}}{\Omega_{SR}}\right)\tau_k^*} \left(\frac{b_1}{b_2}\right)^{\frac{z-s+1}{2}} K_{z-s+1}(2\sqrt{b_1 b_2}).$$
(22)

By putting $\xi = 0$ into (22), the exact OP expression in the case of the pSIC can be achieved in closed-form, as well.

## 4 Numerical Results

In this section, we present theoretical results verified by Monte Carlo simulations to demonstrate the accuracy of the EH MIMO-NOMA network's OP analysis. For this, in all theoretical and simulation results, unless otherwise stated, we assume $K = 3$, the users' power factors and their threshold SINR values are adjusted as $(a_1, \gamma_{th_1}) = (0.6, 1.4)$ $(a_2, \gamma_{th_2}) = (0.3, 2.2)$ and $(a_3, \gamma_{th_3}) = (0.1, 2.5)$, $\Omega_{SR} = (d_{SR})^{-\alpha}$ and $\Omega_{RU} = (d_{RU})^{-\alpha}$, where $d_{SR}$ and $d_{RU}$ are the normalized distance between the BS and relay, the relay and users, respectively and $\alpha$ is the path loss exponent. Here, $d_{SR} = 1 - d_{RU} = 0.5$ and $\alpha = 2$ unless otherwise stated.

Fig. 2 plots the OP versus SNR with different channel conditions and antenna configurations $(m_{SR}, m_{RU}; N_S, N_{Rr}, N_{Rt}, N_U)$. Here, $w = 0.5$, $\zeta = 0.8$ and $\xi = 0$ are assumed. As expected, the analytical and simulation results match perfectly. Besides, the OP performance enhances as $N_S$ and/or $N_{Rr}$ and/or $N_U$ increases and channel conditions $(m_{SR}, m_{RU})$ improve. Specifically, OP $_1$ reduces considerably when the numbers of the antennas of the first and/or second hops $(N_S, N_{Rr}, N_U)$ increase while OP $_2$ and OP $_3$ improve when the number of the antennas of the first hop $(N_S, N_{Rr})$ increase rather than when the number of the receive antennas of the second hop $(N_U)$ increase. For ease exposition, Table 1 and Table 2 are given in order to provide, respectively the values of the SNR that required to achieve OP = 10 $^{-3}$ and the SNR gain advantage of different parameters of $(m_{SR}, m_{RU}; N_S, N_{Rr}, N_{Rt}, N_U)$ for each user. Precisely, under the condition of $(m_{SR}, m_{RU}) = (1,1)$ to achieve OP $_1$ = $10^{-3}$, 9 dB, 11 dB and 6.5 dB SNR gain advantages are obtained for $(N_S, N_{Rr}, N_{Rt}, N_U) = (2,1,2,1)$ over (1,1,2,1), (2,1,2,2) over (2,1,2,1), and (2,2,2,2) over (2,1,2,2), respectively. However, to achieve OP $_2$ = OP $_3$ = $10^{-3}$, 5 dB, 0.5 dB and 8 dB SNR gain advantages are observed for (2,1,2,1) over (1,1,2,1), (2,1,2,2) over (2,1,2,1), and (2,2,2,2) over (2,1,2,2), respectively. On the other hand, under the condition of $(N_S, N_{Rr}, N_{Rt}, N_U) = (2,2,2,2)$, to achieve OP $_1$ = OP $_3$ = OP $_3$ = $10^{-3}$, we observe 5 dB 3.5 dB and 3 dB SNR gain advantages for $(m_{SR}, m_{RU}) = (2,2)$ over (1,1), respectively.

Next, to demonstrate the effect of ipSIC on the OP performance, Fig. 3 shows the OP versus SNR with different values of $\xi$ and antenna configurations $(N_S, N_{Rr}, N_{Rt}, N_U)$. Here, $(m_{SR}, m_{RU}) = (1,1)$, $w = 0.5$, $\zeta = 0.8$ are assumed. It is noticed that the OP $_2$ and OP $_3$ increase when $\xi$ increases. In other words, the OP performance for $U_2$ and $U_3$ deteriorates in the case of ipSIC $(\xi = 0.02)$ rather that in the case of pSIC $(\xi = 0)$. However, OP $_1$ is not affected by ipSIC since $U_1$ does not require the SIC process to detect its signal.

Fig. 4 depicts the OP versus power division ratio $w$ with different values of $\xi$. Here, $(m_{SR}, m_{RU}; N_S, N_{Rr}, N_{Rt}, N_U) = (1,1;2,2,2,2)$, SNR = 20 dB, $\zeta = 0.8$ are assumed. As clearly seen, in both pSIC and ipSIC, the optimum power division ratios for the three users are different from each other. The reason behind this phenomenon is that each user experiences different channel conditions and accordingly the user that has better channel conditions will have less transmission power. For instance, since the $U_3$ has better channel condition than the other users, the transmission from the relay to the $U_3$ can be performed with less power than the others. Hence, $U_3$ takes a smaller value which is about $w_3^{opt} = 0.25$. On the other hand, $U_1$ has the worst channel condition than the other users, the transmission from the relay to the $U_1$ can be done with high power than the others, which means that $U_1$ takes a higher value and from the graph, it is approximate $w_1^{opt} = 0.55$. Finally, $U_2$ is in between other users and hence it takes about $w_2^{opt} = 0.35$. It is notable to mention that the optimum power division ratios for the three users do not affect ipSIC and this is maybe due to that the SIC process is done at the users and any error results only affect the OP, not the optimum power division rations at the EH relay.

Fig. 5 presents the OP performance versus the normalized distance $d_{SR}$ with different values of $\xi$. Here, $(m_{SR}, m_{RU}; N_S, N_{Rr}, N_{Rt}, N_U) = (1,1;2,2,2,2)$, SNR = 10 dB, $w = 0.5$, $\zeta = 0.8$ are assumed. It is obvious that in both pSIC and ipSIC, for all users, the OP performance becomes better when the EH relay is located near to the BS. On the other hand, the OP becomes worse and finally converges to the same value for all users when the EH relay is located far away BS (too much near to the users). One of the reasons for this is that the energy harvested in the relay decreases with the bad condition of the channel between the BS and the relay. Another reason is that the signal transmitted to the users declines due to the bad channel. As a result, the optimal relay location has to be closer to the BS.

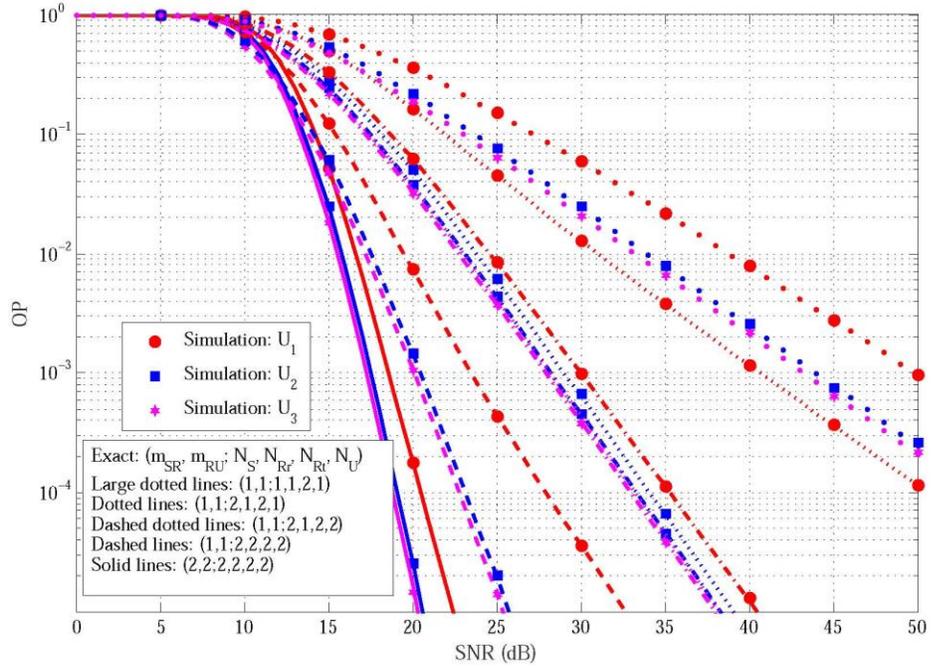

Figure 2: The OP versus SNR in the case of pSIC with different antenna configurations, channel conditions, $w = 0.5$ and $\zeta = 0.8$

Table 1: Values of SNR (in dB) that required to achieve OP = $10^{-3}$ for each user with different parameters of $(m_{SR}, m_{RU}; N_S, N_{Rr}, N_{Rt}, N_U)$.

| $(m_{SR}, m_{RU}; N_S, N_{Rr}, N_{Rt}, N_U)$ | $U_1$ | $U_2$ | $U_3$ |
|---|---|---|---|
| (1,1; 1,1,2,1) | 50 | 4.5 | 44 |
| (1,1; 2,1,2,1) | 41 | 9 | 28.5 |
| (1,1; 2,1,2,2) | 30 | 8.5 | 28 |
| (1,1; 2,2,2,2) | 23.5 | 21 | 20 |
| (2,2; 2,2,2,2) | 18.5 | 7.5 | 17 |

Table 2: SNR gain advantage of different parameters of $(m_{SR}, m_{RU}; N_S, N_{Rr}, N_{Rt}, N_U)$ for each user.

| SNR gain advantage of (in dB) | $U_1$ | $U_2$ | $U_3$ |
|---|---|---|---|
| (1,1; 2,1,2,1) over (1,1; 1,1,2,1) | 9 | 15.5 | 15.5 |
| (1,1; 2,1,2,2) over (1,1; 2,1,2,1) | 11 | 0.5 | 0.5 |
| (1,1; 2,2,2,2) over (1,1; 2,1,2,2) | 6.5 | 7.5 | 8 |
| (2,2; 2,2,2,2) over (1,1; 2,2,2,2) | 5 | 3.5 | 3 |

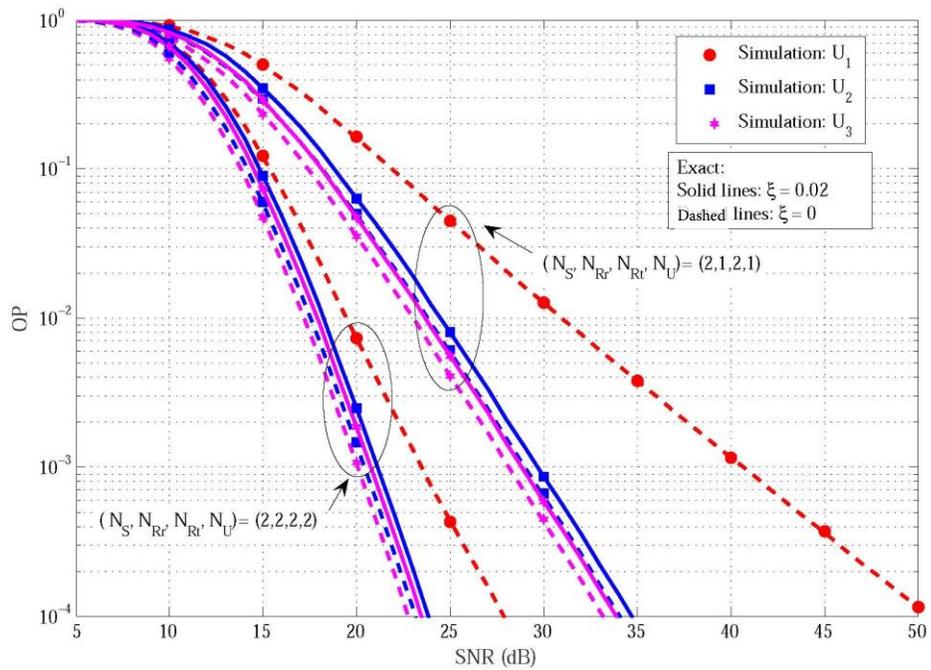

Figure 3: The OP versus SNR in the case of ipSIC with different values of $\xi$, antenna configurations $(N_S, N_{Rr}, N_{Rt}, N_U)$, $(m_{SR}, m_{RU}) = (1,1)$, $w = 0.5$ and $\zeta = 0.8$.

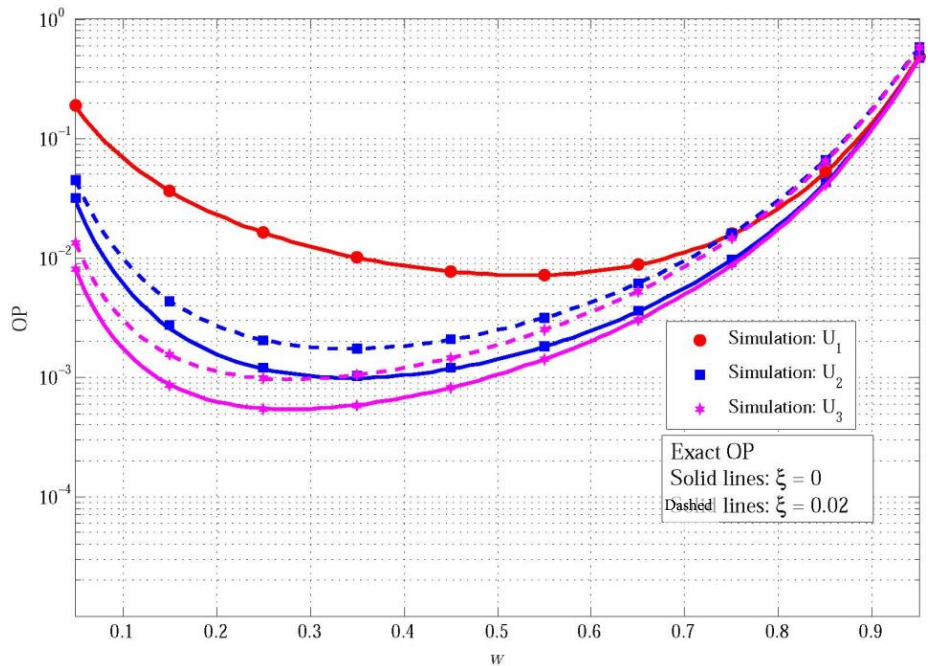

Figure 4: The OP versus $w$ with different values of $\xi$, $(m_{SR}, m_{RU}; N_S, N_{Rr}, N_{Rt}, N_U) = (1,1;2,2,2,2)$, SNR = 20 dB and $\zeta = 0.8$.

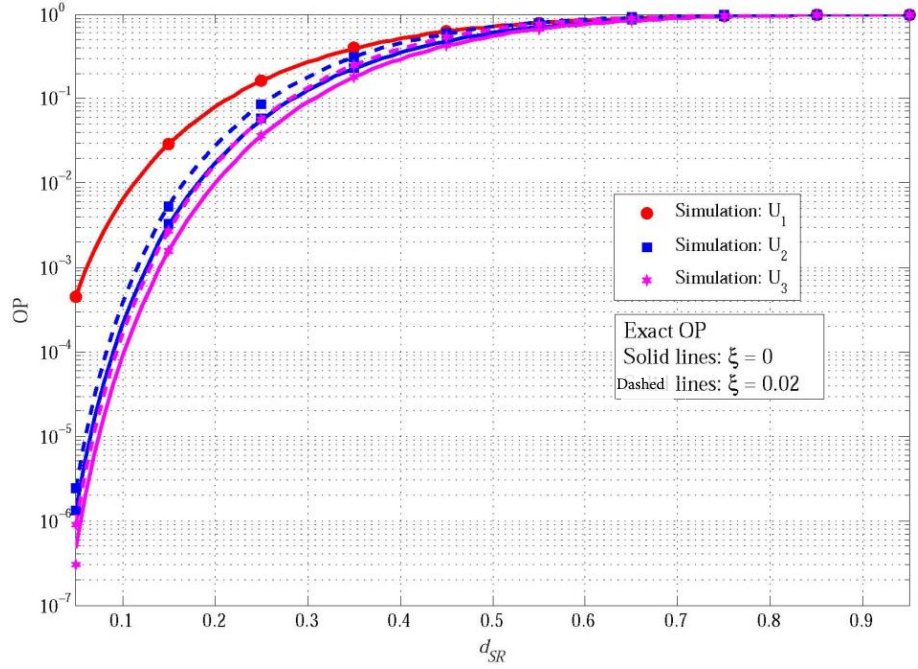

Figure 5: The OP versus $d_{SR}$ with different values of $\xi$, $(m_{SR}, m_{RU}; N_S, N_{Rr}, N_{Rt}, N_U) = (1,1;2,2,2,2)$, SNR = 10 dB, $w = 0.5$ and $\zeta = 0.8$.

## 5  Conclusion

In this paper, the outage behavior of an EH MIMO-NOMA network with the JTRAS scheme is examined in both pSIC and ipSIC cases. The OP analysis is conducted over Nakagami-*m* fading channels and the exact OP expression is derived in closed-form. The theoretical results verified by the Monte Carlo simulation and the results demonstrate that OP performance enhances as the number of the antennas increases and channel conditions. Particularly, the OP performance of the users with bad channel conditions considerably improves as the number of antennas of both hops increases. On the other hand, the OP performance of users with good channel conditions remarkably enhances when the number of the antennas of the first hop increases. Moreover, the OP performance becomes worse in the case of ipSIC rather than that in the case of pSIC. Furthermore, the optimum power division proportions at the EH relay are different for users and the user with good channel conditions will have less transmission power. In addition, the optimal EH relay location has to be closer to the base station and as the EH relay becomes near the users, the OP becomes worse and converges to the same value for all users.

# References

1. L. Dai *et al.,* Non-orthogonal multiple access for 5G: solutions, challenges, opportunities, and future research trends, *IEEE Commun. Mag.*, vol. 53, no. 9, pp. 74-81, Sept. 2015.
2. M. Aldababsa *et al.,* A Tutorial on Non-Orthogonal Multiple Access (NOMA) for 5G and Beyond, *Wirel. Commun. Mob. Comput.*, vol. 2018, pp. 1-24, June 2018.
3. J. Men and J. Ge, Non-Orthogonal Multiple Access for Multiple-Antenna Relaying Networks, *IEEE Commun. Lett.*, vol. 19, no. 10, pp. 1686-1689, Oct. 2015.
4. M. Aldababsa and O. Kucur, Outage performance of NOMA with TAS/MRC in dual hop AF relaying networks, *2017 Adv. Wirel. Opt. Commun. (RTUWO)*, Riga, 2017, pp. 137-141.
5. Y. Zhang *et al.,* Performance Analysis of Nonorthogonal Multiple Access for Downlink Networks With Antenna Selection Over Nakagami-*m* Fading Channels, *IEEE Trans. Veh. Technol.*, vol. 66, no. 11, pp. 10590-10594, Nov. 2017.
6. M. Aldababsa and O. Kucur, Outage and ergodic sumâ€‐rate performance of cooperative MIMO-NOMA with imperfect CSI and SIC, *Int. J. Commun. Syst.*, vol. 33, no. 11, pp. e4405, July 2020.
7. M. Aldababsa and O. Kucur, Cooperative NOMA with Antenna Selection Schemes, *IEEE 2019 27th Signal Process. Commun. Appl. Conf. (SIU)*, Sivas, Turkey, 2019, pp. 1-4.
8. M. Aldababsa and O. Kucur, Performance of cooperative multiple-input multiple-output NOMA in Nakagami-*m* fading channels with channel estimation errors, *IET Commun.*, vol. 14, no. 2, pp. 274-281, Jan. 2020.
9. M. Aldababsa *et al.,* Unified Performance Analysis of Antenna Selection Schemes for Cooperative MIMO-NOMA with Practical Impairments, submitted to *IEEE Trans. Wirel. Commun.*
10. J. Li *et al.,* Performance Study of Cooperative Non-orthogonal Multiple Access with Energy Harvesting, *2019 2nd Int. Conf. Commun. Eng. Technol. (ICCET)*, Nagoya, Japan, 2019, pp. 30-34.
11. M. Hu *et al.,* Ergodic Sum Rate Analysis for Non-Orthogonal Multiple Access Relaying Networks with Energy Harvesting, *2018 2nd IEEE Adv. Inf. Manag., Commun., Electron. Autom. Control Conf. (IMCEC)*, Xi'an, 2018, pp. 474-477.
12. N. Dahi and N. Hamdi, Outage Performance in Cooperative NOMA Systems with Energy Harvesting in Nakagami-*m* Fading, *2018 7th Int. Conf. Commun. Netw. (ComNet)*, Hammamet, Tunisia, 2018, pp. 1-4.
13. P. N. Son and H. Y. Kong, Co-channel interference energy harvesting for decode-and-forward relaying, *Wirel. Pers. Commun.*, vol. 95, no. 4, pp. 3629-3652, Aug 2017.
14. P. N. Son and H. Y. Kong, Energy-Harvesting Decode-and-Forward Relaying under Hardware Impairments, *Wirel. Pers. Commun.*, vol. 96, no. 2, pp. 6381-6395, June 2017.
15. T. A. Le and H. Y. Kong, Performance analysis of downlink NOMA-EH relaying network in the presence of residual transmit RF hardware impairments, *Wirel. Netw.*, vol. 26, no. 2, pp. 1045-1055, Feb. 2020.
16. T.A. Le and H.Y. Kong, Effects of Hardware Impairment on the Cooperative NOMA EH Relaying Network Over Nakagami-*m* Fading Channels, *Wirel. Pers. Commun.*, Nov. 2020.

17. P. N. Son *et al.,* Exact outage analysis of a decode-and-forward cooperative communication network with Nth-best energy harvesting relay selection, *Ann. Telecommun.*, vol. 71, no. 5, pp. 251-263, Jun. 2016.
18. V. P. Tuan and H. Y. Kong, Impact of residual transmit RF impairments on energy harvesting relay selection systems, *Int. J. Electron.*, vol. 104, no. 6, Feb. 2017.
19. W. Guo and W. Guo, Non-Orthogonal Multiple Access Networks with Energy Harvesting and Cooperative Communication, *2018 5th Int. Conf. Inf. Sci. Control Eng. (ICISCE)*, Zhengzhou, 2018, pp. 1163-1167.
20. L. T. Dung *et al.,* Analysis of partial relay selection in NOMA systems with RF energy harvesting, *2018 2nd Int. Conf. Adv. Signal Process., Telecommun. & Computing (SigTelCom)*, Ho Chi Minh City, 2018, pp. 13-18.
21. S. Rao, Relay Selection for Energy Harvesting Cooperative NOMA, *2019 IEEE Int. Conf. Adv. Netw. Telecommun. Syst. (ANTS)*, Goa, India, 2019, pp. 1-6.
22. W. Han *et al.,* Performance analysis for NOMA energy harvesting relaying networks with transmit antenna selection and maximal-ratio combining over Nakagami-*m* fading, *IET Commun.*, vol. 10, no. 18, pp. 2687-2693, 15 12 2016.
23. Q. Wang *et al.,* Performance analysis of NOMA for multiple-antenna relaying networks with energy harvesting over Nakagami-*m* fading channels, *2017 IEEE/CIC Int. Conf. Commun. in China (ICCC)*, Qingdao, 2017, pp. 1-5.
24. B. Demirkol and O. Kucur, Outage Performance of Antenna Selection Schemes in NOMA Networks using Amplify-and-Forward Energy Harvesting Relay, *2020 28th Signal Process. Commun. Appl. Conf. (SIU)*, Gaziantep, Turkey, 2020, pp. 1-4.
25. Y. Yu *et al.,* Antenna Selection for MIMO Nonorthogonal Multiple Access Systems, *IEEE Trans. Veh. Technol.*, vol. 67, no. 4, pp. 3158-3171, April 2018.
26. Q. Li *et al.,* Joint antenna selection for MIMO-NOMA networks over Nakagami-*m* fading channels, *2017 IEEE/CIC Int. Conf. Commun. in China (ICCC)*, Qingdao, 2017, pp. 1-6.
27. Y. Yu *et al.,* Antenna Selection in MIMO Cognitive Radio-Inspired NOMA Systems, *IEEE Commun. Lett.*, vol. 21, no. 12, pp. 2658-2661, Dec. 2017.
28. M. Aldababsa and O. Kucur, Outage Performance of NOMA with Majority Based TAS/MRC Scheme in Rayleigh Fading Channels, *IEEE 2019 27th Signal Process. Commun. Appl. Conf. (SIU)*, Sivas, Turkey, 2019, pp. 1-4.
29. M. Aldababsa and O. Kucur, Majority Based TAS/MRC Scheme in Downlink NOMA Network with Channel Estimation Errors and Feedback Delay, *arXiv:1903.07922*, 2019.
30. M. Aldababsa and O. Kucur, Majority based antenna selection schemes in downlink NOMA network with channel estimation errors and feedback delay, *IET Commun.*, vol. 14, no. 17, pp. 2931-2943, Oct. 2020.
31. M. Aldababsa and O. Kucur, BER Performance of NOMA Network with Majority Based JTRAS Scheme in Practical Impairments, *Int. J. Electron. Commun. (AEÜ)*, to be published.
32. A. Nasir *et al.,* Relaying Protocols for Wireless Energy Harvesting and Information Processing, *IEEE Trans. Wirel. Commun.*, vol. 12, no. 7, pp. 3622-3636, July 2013.
33. Han, T. Lv and X. Zhang, Outage Performance of NOMA-based UAV-Assisted Communication with Imperfect SIC, *2019 IEEE Wirel. Commun. Netw. Conf. (WCNC)*, Marrakesh, Morocco, pp. 1-6, 2019.

34. Yılmaz, and O. Kucur, Unified analysis of transmit, receive and hybrid diversity techniques over generalized-K channels, *Wirel. Commun. Mob. Comput.*, vol. 16, no. 13, pp. 1798- 1808, 2016.
35. IS. Gradshteyn and IM. Ryzhik, *Table of Integrals, Series, and Products*, 7th Ed. Academic Press: San Diego, CA, 2007.